\documentclass[superscriptaddress,showpacs,preprintnumbers,amsmath,amssymb,aps,pre]{revtex4}
\usepackage{graphicx}
\usepackage{dcolumn}
\usepackage{bm}

\begin{document}
\title{Additional information\footnote{Submitted to Physical Review E on October, 21, 2005} for the paper `Entropy of seismic electric signals: Analysis in natural time under time-reversal'}
\author{P. A. Varotsos}
\email{pvaro@otenet.gr}
\affiliation{Solid State Section, Physics Department, University of Athens, Panepistimiopolis, Zografos 157 84,
Athens, Greece}
\affiliation{Solid Earth Physics Institute, Physics Department, University of Athens, Panepistimiopolis, Zografos 157 84, Athens, Greece}
\author{N. V. Sarlis}
\affiliation{Solid State Section, Physics Department, University of Athens, Panepistimiopolis, Zografos 157 84,
Athens, Greece}
\author{E. S. Skordas}
\affiliation{Solid State Section, Physics Department, University of Athens, Panepistimiopolis, Zografos 157 84,
Athens, Greece}
\affiliation{Solid Earth Physics Institute, Physics Department, University of Athens, Panepistimiopolis, Zografos 157 84, Athens, Greece}
\author{H. K. Tanaka}
\affiliation{Earthquake Prediction Research Center, Tokai University 3-20-1, Shimizu-Orido, Shizuoka 424-8610, Japan}
\author{M. S. Lazaridou}
\affiliation{Solid State Section, Physics Department, University of Athens, Panepistimiopolis, Zografos 157 84,
Athens, Greece}

\begin{abstract}
After the submission of the paper, three strong earthquakes with magnitude around 6.0-units
occurred on October 17 and October 20, 2005, with epicenters in the Aegean Sea, at a distance {\em only}
100km from MYT station at which the intense signals $M_1$ to  $M_4$ -analyzed in the main
text- have been recorded. This confirms experimentally the proposed criterion we used for
 the classification of these signals as Seismic Electric Signals (SES). Moreover,
 we show that, if we follow the procedure described in [P.A. Varotsos, N. V. Sarlis,
 H. K. Tanaka and E. S. Skordas {\it Phys. Rev. E} {\bf 72}, 041103 (2005)], the analysis
 in the natural time of the seismicity after the SES initiation allows the estimation of the
 time window of the impending earthquakes with very good accuracy.
\end{abstract}
 \maketitle
\begin{figure}
\includegraphics{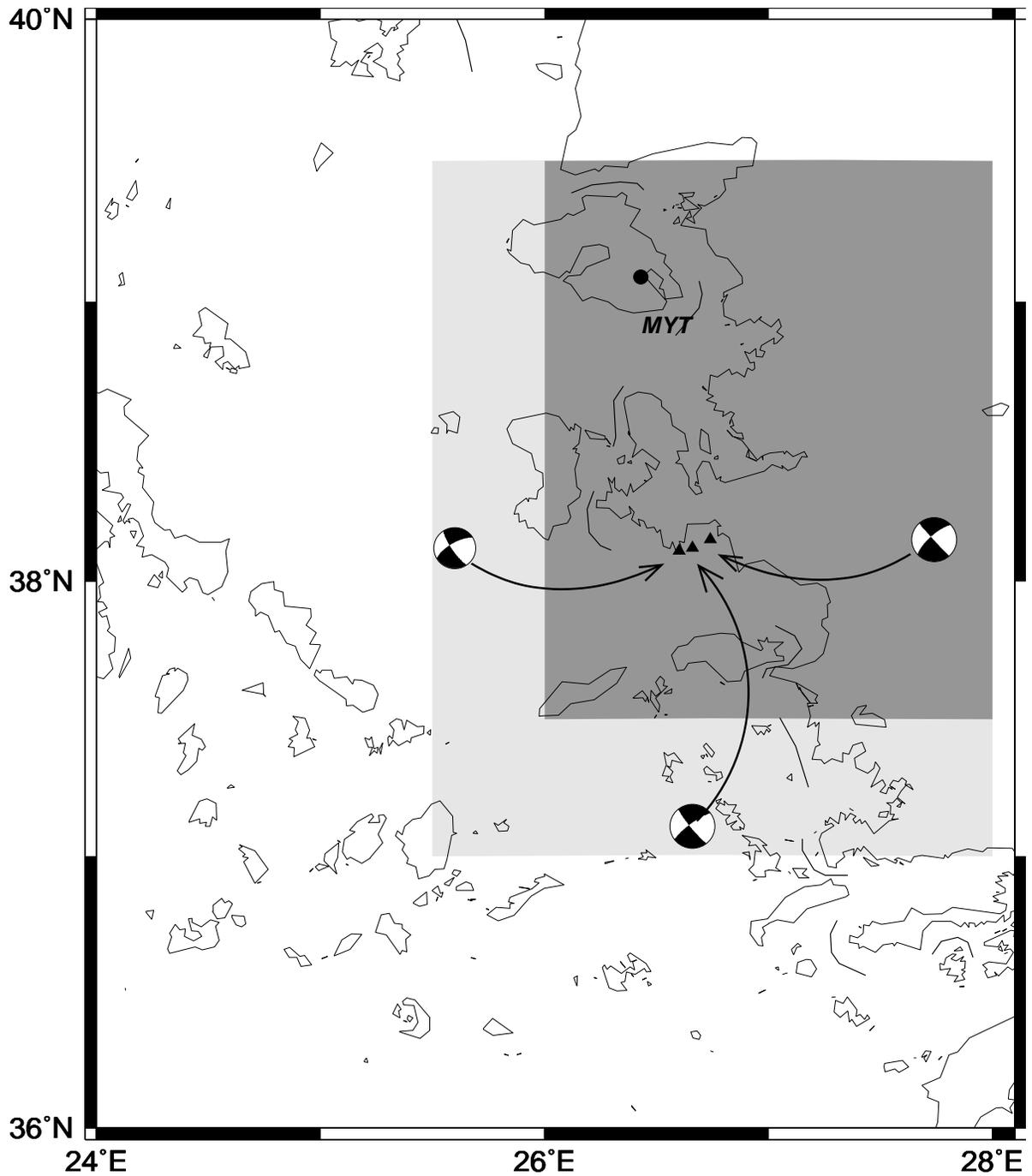}
\caption{Map of the area surrounding the  measuring station of MYT and the epicenters of the three strong EQs that occurred on October 17 and October 20,2005.  The earthquake
mechanisms of all three EQs are also shown. The seismicity subsequent to the SES initiation has been  studied in the gray shaded regions.} \label{f1}
\end{figure}

On October 17, 2005 two strong earthquakes (EQs) with magnitude around 6.0-units occurred at
05:45:20 and 12:46:57 UT with epicenters (according to USGS) at 38.15$^oN$,26.68$^oN$ and
 38.13$^oN$,26.65$^oN$, respectively (see Fig.\ref{f1}).  At the same epicenter, a third almost equally strong earthquake
 occurred at 21:40 UT on October 20,2005. All the three epicenters lie at a distance of around
 100km from Lesvos island, at which the MYT station -on the dipoles of which the intense
 signals $M_1$ to  $M_4$ (Fig.1(a),(b) of the main text) have been recorded- is located.
 This verifes that these signals are actually Seismic Electric Signals (SES) as classified in advance (i.e., upon the initial submission of the paper on April 16,2005) in the main text on the basis of the entropy criterion proposed.

\begin{figure}
\includegraphics{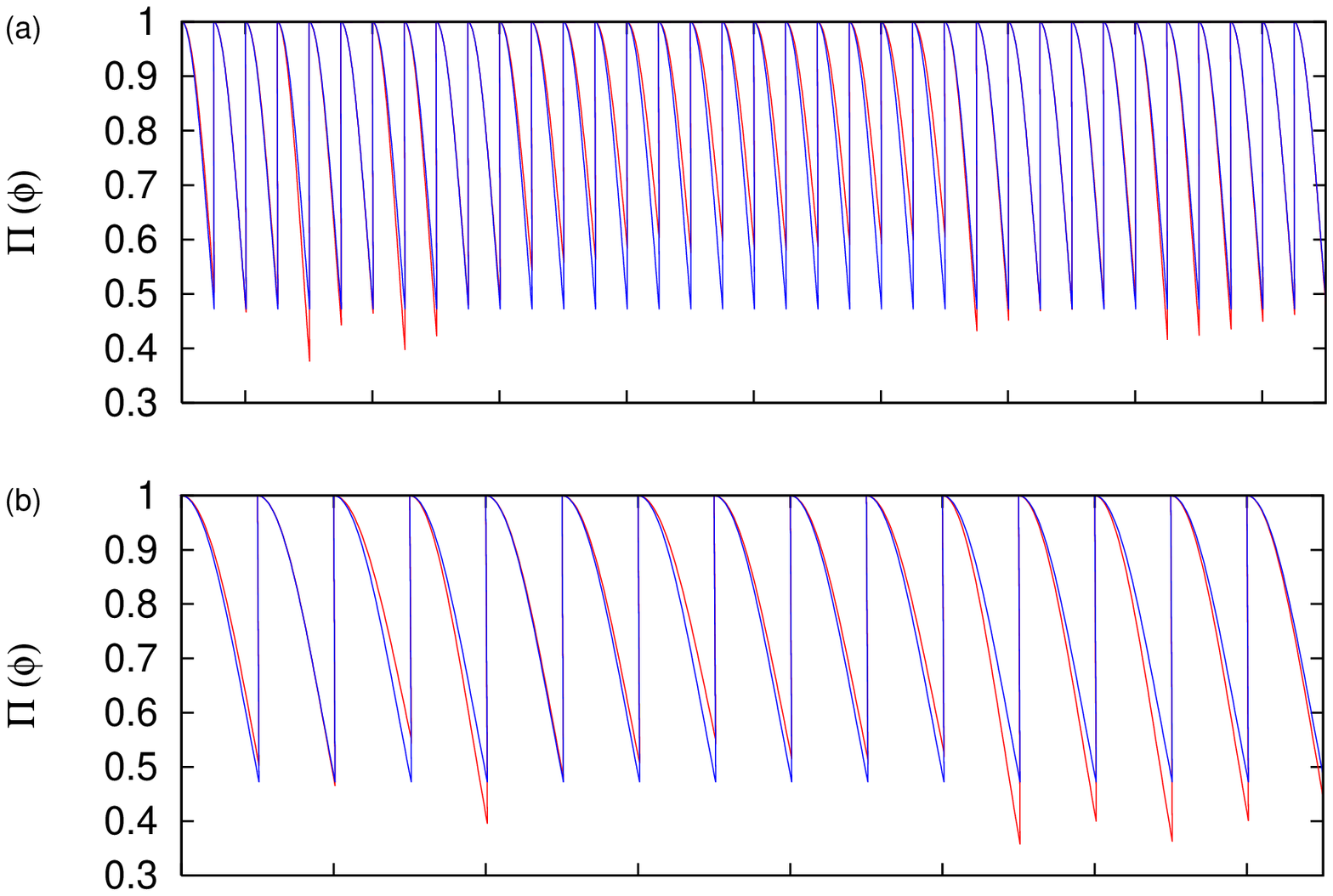}
\caption{The normalized power spectrum(red) $\Pi (\phi )$  of the seismicity in area A as it evolves event by event after the
initiation of the SES activities $M_1$ to  $M_4$. The two examples presented correspond to the two different magnitude thresholds 3.4 and 3.6 in (a),(b) respectively. In each case only the region $\phi \in [0,0.5]$ is depicted (for the reasons
discussed in Refs.\cite{NAT01,prlop}), whereas the  $\Pi (\phi )$ of Eq.(\ref{fasma}) is depicted by blue color.   }
\label{f2}
\end{figure}

\begin{figure}
\includegraphics{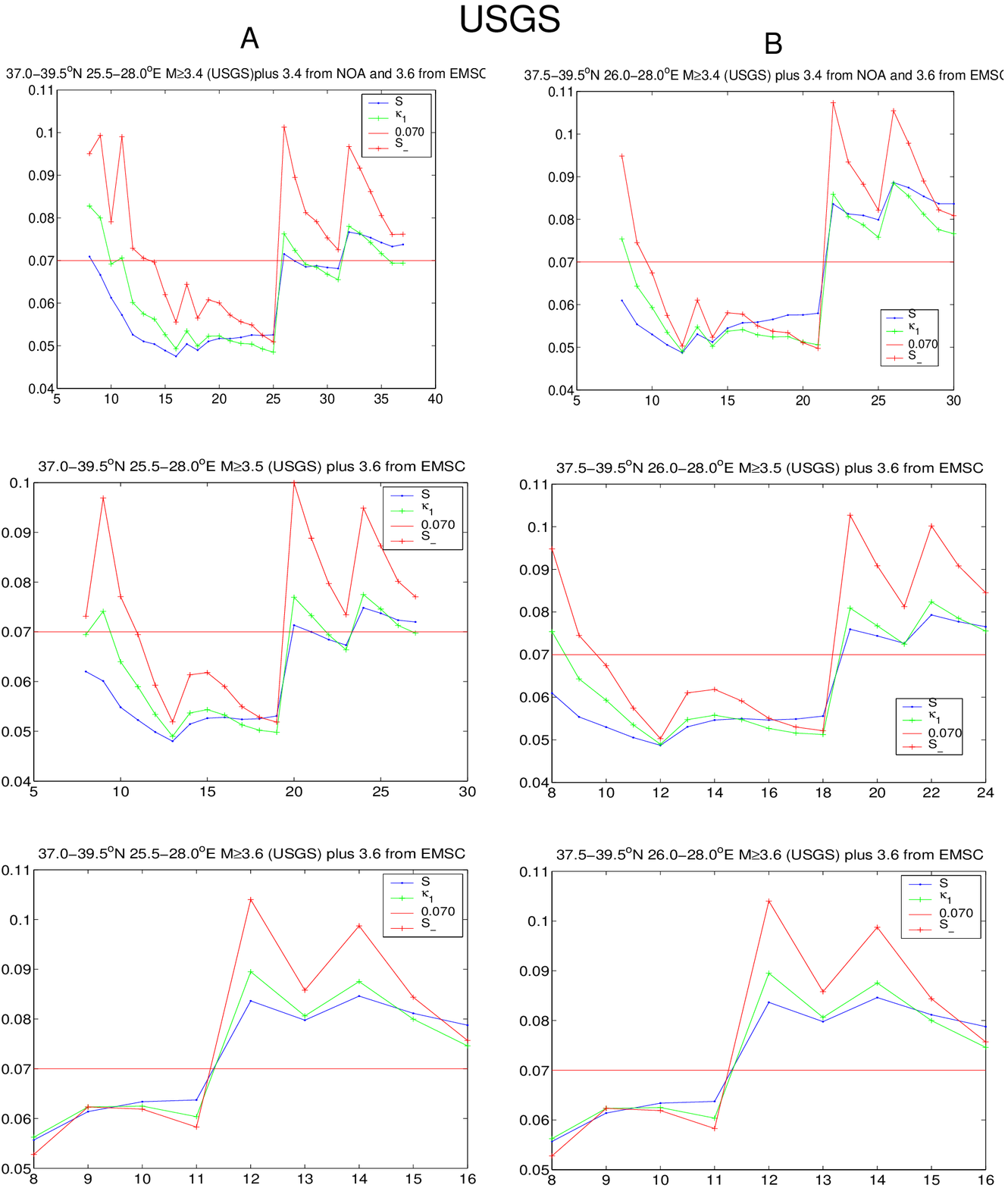}
\caption{Evolution of  $\kappa_1$, $S$ and $S_{-}$ upon using the USGS catalogue for various
magnitude ($M_L$ or $M_D$) thresholds for the two regions A and B.}
\label{f4}
\end{figure}

We now follow the procedure described in Refs.\cite{NAT01,prlop}, in order to investigate whether the time window of the impending strong EQs could have been estimated. We consider either the region A:$N_{37.0}^{39.5}E_{25.5}^{28.0}$ or
 the region B:$N_{37.5}^{39.5}E_{26.0}^{28.0}$, which
 surround the EQ epicenters and the MYT station (see Fig.\ref{f1}), and study how the seismicity evolved after the SES initiation.
 If we set the natural time for seismicity zero at the initiation of the concerned SES activities, we form
 time series of seismic events in natural time for various time windows as the number $N$ of consecutive (small)
 EQs increases. We now compute the normalized power spectrum \cite{NAT01,prlop} in natural time $\Pi (\phi )$ for
 each of the time windows and the results are depicted in Fig.\ref{f2}.  As examples
 we consider in this figure  two magnitude thresholds (herafter referring to the local magnitude $M_L$ or
 the `duration' magnitude $M_D$) 3.4 (upper) and 3.6 (lower). In the same figure, we plot in blue the power spectrum obeying
 the relation
\begin{equation}
\Pi ( \omega ) = \frac{18}{5 \omega^2}
-\frac{6 \cos \omega}{5 \omega^2}
-\frac{12 \sin \omega}{5 \omega^3}
\label{fasma}
\end{equation}
which holds\cite{NAT01,NAT02,NAT02A} when the system enters into the {\em critical} stage ($\omega = 2\pi \phi$, where $\phi$ stands for the natural frequency\cite{NAT01,NAT02,newbook}).
An inspection of Fig.\ref{f2} reveals that the red line approaches the blue line as $N$ increases and a {\em coincidence} occurs at the last small
event which had a magnitude 3.6 and occurred at 04:31 UT on October 17, 2005, i.e., roughly one hour before the first strong EQ. To ensure that this coincidence is a {\em true} one\cite{NAT01,prlop,newbook} we also calculate the evolution of $\kappa_1$,$S$ and $S_{-}$ (cf.
$\kappa_1$ stands for the variance $\kappa_1\equiv \langle \chi^2 \rangle -\langle \chi \rangle^2$ as explained in Refs.\cite{NAT01,NAT02}) and the results are depicted in Fig.\ref{f4} for three
magnitude thresholds 3.4, 3.5 and 3.6.

We now further comment on the aforementioned results. Since the strong EQs occurred in the
border between Greece and Turkey, the seismicity catalogues of neither Greek nor the
Turkish Institutes can be considered as complete for small magnitudes. Hence, we preferred
here to make the calculations by using the United States Geological Survey (USGS) catalogue (see Table \ref{tab1}).
Irrespective if we use the seismicity in the region $N_{37.0}^{39.5}E_{25.5}^{28.0}$ or
in  the smaller region $N_{37.5}^{39.5}E_{26.0}^{28.0}$, the coincidence occurs upon the
occurence of the aforementioned  3.6 EQ (almost 1 hour before the first strong EQ). The
magnitude of this EQ comes from the European-Mediterranean Seismological Centre (EMSC) (see the
corresponding announcement in Fig.\ref{f3}) since it has not yet been reported by USGS. Note that
if we take the magnitude of this EQ to be somewhat larger, then the first box in Fig.\ref{f4} (which has been plotted for magnitude threshold 3.4) shows that the coincidence occurs on the last but one event, i.e., on October 13,2005 (almost three days before the first strong EQ).

The following comment might be useful:
  In the analysis of signals depicted in Figs1(a),(b) we proceeded
as follows: We first analyzed $M_1$ (recorded on March 21, see Fig1(a))
and found that it obeys the criterion (i.e., $S$ and $S_{-}$ smaller than $S_u$). We then turned to  the recordings on March 23: we first considered $M_4$ -which is well
distinguishable from the others in view of its opposite polarity- and
found again that the criterion is fullfilled. As for the remaining recordings of March 23(comprising $M_2$ and $M_3$) we checked both possibilities that is: (1) we
considered $M_2$ and $M_3$ together and found that not only the criterion
was violated but also the $M_2+M_3$ signal (altogether) behaved like signals obtained
from a "uniform" distribution(i.e., $\kappa_1$ was around $1/12$). (2)on the
other hand, if we analyse separately $M_2$ and $M_3$ then both results obeyed
the criterion. (thus being consistent with the conclusions drawn from $M_1$ and $M_4$). Hence, we preferred to draw in Fig.1b the latter possibility.
    It has been earlier suggested (see pages 18 and 114 of  Ref.[10] of
the main text) that the change of the SES polarity might be associated
with a different EQ source mechanism. The first three SES activities $M_1$,
$M_2$ and $M_3$ -which have been plotted in Figs. 1(a,b) do have the same
polarity and hence- might be associated with the 3 EQs that have already
occurred having the {\em same} mechanism. If this suggestion is correct, the last signal, i.e., $M_4$, with opposite polarity, should correspond to another EQ with different
mechanism which has not occurred yet.

\begin{figure}
\includegraphics{Fig4}
\caption{The detailed European-Mediterranean Seismological Centre (EMSC) report for the 3.6 EQ that occured almost one hour before the first strong EQ.}
\label{f3}
\end{figure}


\begin{table}
\caption{The United States Geological Survey (USGS) Catalogue for the area A under discussion together with the last event reported
by EMSC (see Fig.\ref{f3}).}  \label{tab1}
\includegraphics{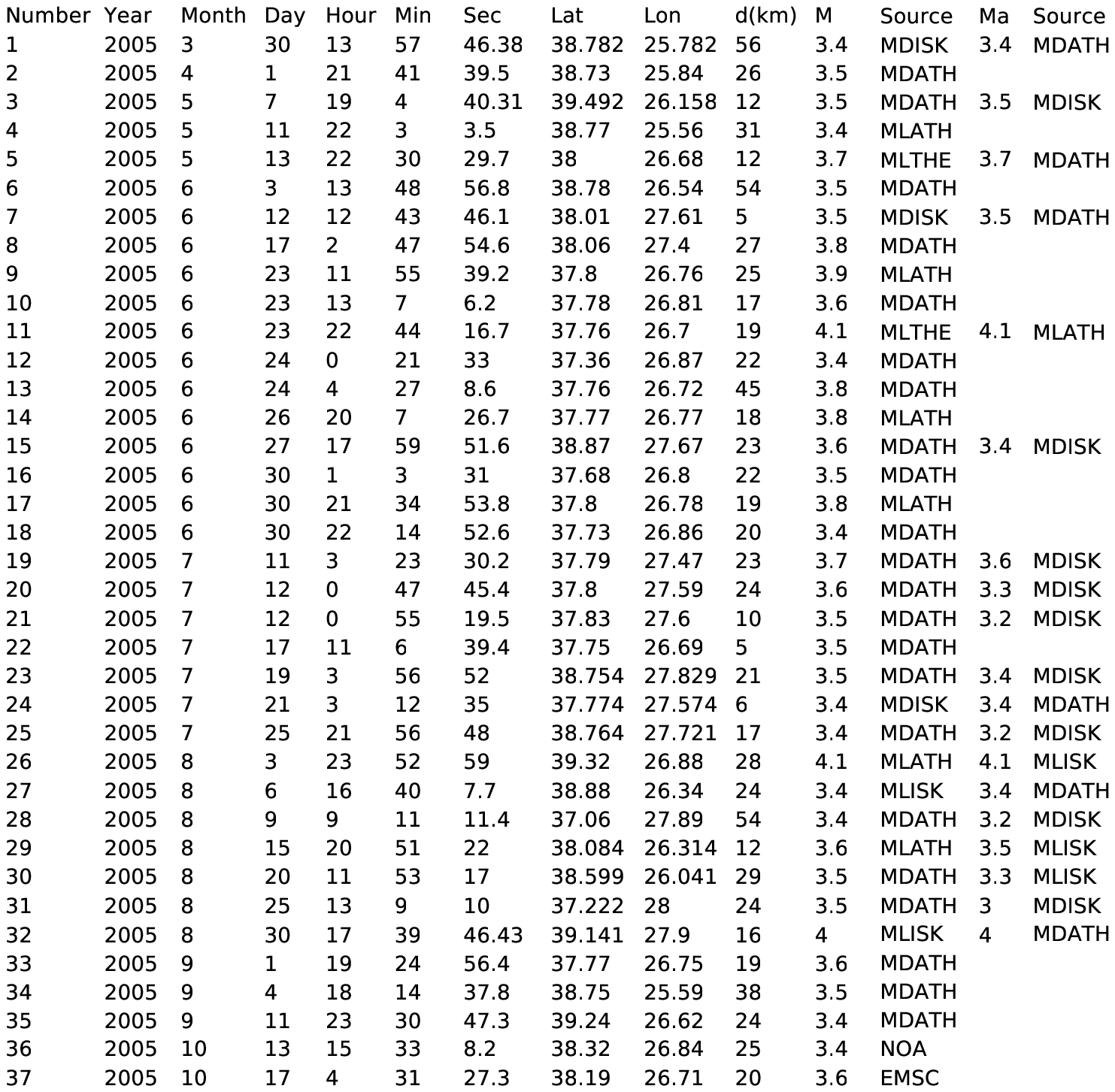}
\end{table}




\bibliographystyle{apsrev}

\end{document}